\documentclass[twocolumn,showpacs,pra,preprintnumbers,amsmath,amssymb]{revtex4}
\usepackage{graphicx}% Include figure files
\usepackage{dcolumn}% Align table columns on decimal point
\usepackage{bm}% bold math

\begin{document}

%\preprint{APS/123-QED}

\title{Non-Markovian entanglement dynamics of quantum continuous
  variable systems in thermal environments}

\author{Kuan-Liang Liu}

\author{Hsi-Sheng Goan}
\email{goan@phys.ntu.edu.tw}
\affiliation{Department of Physics and Center for Theoretical
  Sciences, National Taiwan University, and\\
National Center for Theoretical Sciences, Taipei 10617, Taiwan}

\date{\today}% It is always \today, today,

\begin{abstract}

We study two continuous variable systems (or two harmonic
oscillators) and investigate their entanglement evolution under the
influence of non-Markovian thermal environments. The continuous
variable systems could be two modes of electromagnetic fields or two
nanomechanical oscillators in the quantum domain. We use quantum
open system method to derive the non-Markovian master equations of
the reduced density matrix for two different but related models of
the continuous variable systems. The two models both consist of two
interacting harmonic oscillators. In model A, each of the two
oscillators is coupled to its own
independent thermal reservoir, while in model B the two oscillators are 
coupled to a common reservoir. To quantify the degrees of entanglement
for the bipartite continuous variable systems in Gaussian states, logarithmic
negativity is used. We find that the dynamics of the quantum
entanglement is sensitive to the initial states, the
oscillator-oscillator interaction, the oscillator-environment
interaction and the coupling to a common bath or to different,
independent baths.

\end{abstract}


\maketitle

\section{Introduction}

Quantum information and computation is a nascent and interdisciplinary
field that exploits
quantum effects to
compute and process information in ways that are faster or more
efficient than or even impossible on conventional computers or
information processing devices.  This field has typically concerned
itself with the manipulations of discrete systems such as
quantum bits (qubits).
Recently, the extension to continue variables such as position,
momentum, or the quadrature amplitudes of electromagnetic fields
has led to the illuminating concept of continuous variable quantum information
processing \cite{Braunstein05}.
This includes the experimental realization of quantum
teleportation \cite{Furusawa98,Zhang03,Bowen03} and
the demonstration of
quantum key distribution \cite{Yuen98,Grosshans03} for continuous
optical fields, and the successful definition of the notion of universal
quantum computation over continuous variables \cite{Braunstein05}.

Advances in current technology have allowed the fabrication of very small
mechanical cantilevers or oscillators with high frequencies, and allowed
their operation and manipulation at very low temperatures
\cite{Roukes,Blencowe04}.
In the regime when the individual mechanical vibration quanta are of the
order of the thermal energy,
the motion of the mechanical oscillators is close to or
on the verge
of the quantum limit.
Recently investigations devoted
to observing quantum effect in the truly
solid-state mechanical oscillators have been reported \cite{Cleland,
  Schwab, Schwab06, Armour, Marshall, Mancini,  Pirandola, Xue07,
  Plenio02, Plenio04, Eisert}.
The Hilbert space of quantized electromagnetic fields is equivalent to
the Hilbert space of the quantum Harmonic oscillators.
Thus, in addition to quantum optics system, it may be possible to
implement continuous variable quantum
information processing in the nanomechanical oscillator systems.
This is particularly interesting in that it provides a stepping
stone towards quantum state control and a platform to explore the
transition from quantum to classical world in mechanical systems
that consist of many-million atoms \cite{Mancini,  Pirandola, Xue07,
Plenio02, Plenio04, Eisert}.

Quantum entanglement has been considered as key resources in many
 aspects and applications of quantum information processing.
%Quite often, it is assumed that entanglement would not diminished or
%degraded during information transmission and processing. However,
In the real world, quantum coherence and entanglement of quantum
systems will inevitably be influenced and degraded by external environments.
There have been several investigations of decoherence
and quantum entanglement of continuous variable systems under open
system dynamics in the literature
\cite{Duan97,Hiroshima01,Scheel01,Wilson03,Olivares03,Jakub,Serafini04,Serafini05,Benatti06,Ban,Dodd04, Dodd04a}.
But in those investigations, Markovian approximation or (and)
rotating-wave approximation (RWA) is (are) assumed.
However, if the short time interval or regime, comparable with the
environment correlation time, is concerned, or if the environments is
structured with a particular spectral density, then the non-Markovian
environmental effect could become significant.
For example, in the case
when a high-speed quantum information processing is required,
the non-Markovian effect becomes important since the
typical characteristic time of the relevant system may be comparable
with the reservoir correlation time.
Besides, when the typical system characteristic time is
comparable with
the decoherence and dissipation times, the other approximation, RWA,
widely used in quantum optics master equation \cite{Gardiner00} to
describe open quantum system may not apply.
This is particularly the case for nanomechanical oscillators (or beams) as
their fundamental vibration
frequency $\Omega$ could currently just reach a few GHz
\cite{Huang}, still much smaller than the optical frequency of
$10^{15}$ Hz.
%To be operated deeply in the quantum regime, the
%nanomechanical oscillators are envisaged to be cooled to a very low
%temperature such that $K_B T \ll \hbar \Omega$.
%In Refs. \cite{Duan97,Hiroshima01,Scheel01,Wilson03,Olivares03,Jakub,
%Serafini04} the degration of entanglement of two-mode Gaussian
%states in thermal baths  or sueezed
%reserviors has been analyzed in terms of
%different entanglement measures but all within the RWA-Markovian
%approximation.
%Recently, Ban \cite{Ban} has discussed decoherence and entanglement
%evolution  under the influence of non-Markovian quantum channels but
%the master equation derived is still under RWA.
%References \cite{Dodd04, Dodd04a} have
%discussed the destruction of entanglement of two-particle systems
%under open system dynamics described by a class of
%Markovian master equations but without making the RWA.
Thus a detailed investigation of the non-Markovian quantum
entanglement dynamics in a more general setting without RWA is
demanding. The main purpose of this paper is to present such a detailed
analysis. The analysis is, as mentioned, of great importance to
quantum nanomechanical oscillator systems and its relevance to quantum
optics systems is also obvious.

\begin{figure}
    \includegraphics[width=8.5cm]{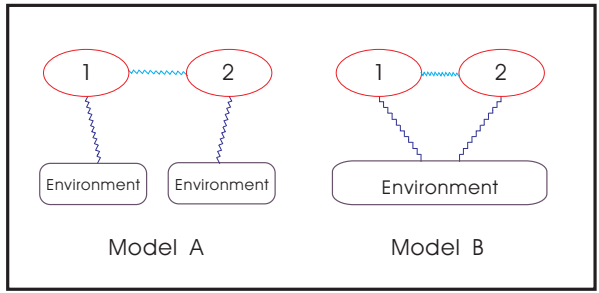}
        \caption{(Color online). The schematic illustration of the two
          models investigated. The two models both consist of two
interacting harmonic oscillators. In model A, each of the two
oscillators is coupled to its own
independent thermal reservoir, while in model B the two oscillators are 
coupled to a common reservoir.}\label{all_models}
\end{figure}

We study, in this paper, two harmonic oscillators
 in the quantum domain and investigate their entanglement evolution
 under the influence of thermal environments \cite{Jakub, Plenio02,
   Plenio04, Eisert, Serafini05, Ban}.
Two different but related models of harmonic oscillator systems are
investigated (see Fig. \ref{all_models}). Model A consists of two
interacting harmonic oscillators, each coupled to its own
independent reservoir. The two oscillators may be envisaged to be sufficiently
spatially apart and may thus be relevant to
 the related setup for applications in quantum communication
and teleportation. Model B also consists of two interacting harmonic
oscillators, but both coupled to a common reservoir. They may be spatially
close and could be useful for possible applications in quantum
computation or other quantum information processing tasks. The
coupling between the two oscillators, and the coupling between the
oscillators and the environments are, if present, all bilinear in
their respective positions (or coordinates). 
In practice,  there may not be direct interaction between two
 optical fields, and this can be simply achieved by setting the
 coupling between the two oscillators to be zero in our models. 
On the other hand, a controllable and tunable
interaction between two nanomechanical oscillators can be introduced
by applying a voltage made from the metallic film fabricated on
their surfaces \cite{ Schwab,Eisert,Goan05}.
Thus the two models
discussed here are applicable to quantum nanomechanical oscillator
 systems and quantum optics systems.

In the context of two modes of electromagnetic field embedded in a
thermal environment, Ref. \cite{Jakub} derived a condition which
states that if the state of the two modes is initially sufficiently
squeezed, it will always remain entangled independently of the
strength of the coupling to the environment.
%Each of the two modes
%has the Hilbert space equivalent to the Hilbert space of the
%Harmonic oscillator. As a result,
The model studied in Ref.
\cite{Jakub} is the same as our model B but without the interaction
between the two oscillators. The conclusion in
Ref. \cite{Jakub} was derived however using the RWA-Markovian master
equation. Here we investigate whether the condition
presented in Ref. \cite{Jakub} is still valid or needs some
modification in the non-Markovian case. We find that in the case of
non-Markovian dynamics the condition depends also on the interaction
strength between the system and environment.

This paper is organized as follows. In Sec. \ref{sec2} the
Hamiltonian and non-Markovian master equations for our two models
are presented. In Sec. \ref{sec42} we introduce the concept of
renormalization and counter term. In Sec. \ref{sec4} we introduce
the logarithmic negativity to quantify entanglement for the two
oscillators in our models. This entanglement monotone of logarithmic
negativity is conveniently computable for general Gaussian states,
and could 
provide a proper quantification of entanglement particularly 
for two-mode Gaussian states. We thus discuss
%, in Sec. \ref{sec3} 
the covariance matrix and two-mode
squeezed vacuum state, a subclass of Gaussian states which we will
use as initial states. In Sec. \ref{sec41},
the covariance matrix evolution equations obtained from the master equations
or the Fokker-Planck equations of Wigner function 
are described.  In Sec.\ \ref{sec5} we present and discuss our
results of entanglement dynamics based on the logarithmic negativity
calculated through the evolution equations of 
the covariance matrix. Finally, we investigate whether the
entanglement survival (or separability time) 
condition under RWA-Markovian approximation in
Ref. \cite{Jakub} is still valid in the non-Markovian case. A
conclusion is given in Sec. \ref{sec6}.

\section{\label{sec2}Hamiltonian and non-Markovian master equations of two models}

%\subsection{The non-Markovian master equations}
In this section we introduce two different but related models
considered in this paper (see Fig. \ref{all_models}) and derive
their corresponding quantum master equations up to the second order
with respect to the system-environment coupling constant. 
The two models both consist of two 
interacting harmonic oscillators. In model A, each of the two
oscillators is coupled to its own
independent thermal reservoir, while the two oscillators are  
coupled to a common thermal reservoir in model B.

The Hamiltonian of the system of interest for the two models
can be written as
\begin{eqnarray}
 H_s=H_{s1}+H_{s2}+V_{12},
\end{eqnarray}
where $H_{s1}$ and $H_{s2}$ are the Hamiltonian of the two
subsystems respectively and $V_{12}$ is the interaction between
them. They can be written as
\begin{eqnarray}
  &H_{s1}=&\frac{p_x^2}{2M_x}+\frac{1}{2}M_x\Omega_x^2x^2\label{sys1_hamiltonian},\\
  &H_{s2}=&\frac{p_y^2}{2M_y}+\frac{1}{2}M_y\Omega_y^2y^2\label{sys2_hamiltonian},\\
  &V_{12}=&\lambda xy,
 \end{eqnarray}
where $M_x$ and $M_y$ are the masses and $\Omega_x$ and $\Omega_y$
are the frequencies of the two subsystems (oscillators),
respectively, and $\lambda$ is the coupling constant between the two
subsystems. To consider the case of two non-interacting oscillators
in, we can simply set $\lambda$ to zero.

We assume that the environments could be described as ensembles of
harmonic oscillators and interact bilinearly through their position
operators with the system. The Hamiltonian of the two independent
environments for Model A is thus
\begin{eqnarray}
  H_{\varepsilon}&=&H_{\varepsilon1}+H_{\varepsilon2},\notag\\
  &=&\sum_n(\frac{{p_n^{(1)}}^2}{2m_n^{(1)}}+\frac{1}{2}m_n^{(1)}{\omega_n^{(1)}}^2{q_n^{(1)}}^2)\notag\\
   &&+\sum_n(\frac{{p_n^{(2)}}^2}{2m_n^{(2)}}+\frac{1}{2}m_n^{(2)}{\omega_n^{(2)}}^2{q_n^{(2)}}^2),
\end{eqnarray}
and the interaction between the two subsystems and reservoirs for
Model A is
\begin{eqnarray}
  V&=&V_{1}+V_{2}\notag\\
  &=&\sum_n\lambda_n^{(1)} q_n^{(1)} x+\sum_n\lambda_n^{(2)} q_n^{(2)} y,
\end{eqnarray}
where $\lambda_n^{(1)}$ and $\lambda_n^{(2)}$ are the coupling
strengths to their own individual reservoir, respectively.

On the other hand, the Hamiltonian of the environment for the two
subsystems coupled to a common reservoir (Model B) is
\begin{eqnarray}
 H_\varepsilon=\sum_n(\frac{p_n^2}{2m_n}+\frac{1}{2}m_n\omega_n^2q_n^2),
\end{eqnarray}
and the coupling between two subsystems and reservoirs for Model B
are
\begin{eqnarray}
  V&=&V_{1}+V_{2}\notag\\
  &=&\sum_n\lambda_n^{(1)} q_n x+\sum_n\lambda_n^{(2)} q_n y.
\end{eqnarray}

Using the perturbative expansion to the second order in the
system-environment coupling strength, we obtain the equation of motion
for the reduced density matrix $\rho(t)$ of the system of interest as \cite{Paz_lecture}
\begin{eqnarray}
\dot{\rho}(t)&=&{\frac{1}{i\hbar}[H_s, \rho(t)]}
 -e^{-\frac{i}{\hbar}H_st} \notag\\
%\big(\frac{1}{i\hbar}\mbox{Tr}_\varepsilon
%[\tilde{V}(t),\tilde{\rho}_\tau(0)]}\notag\\
&&\hspace{-0.6cm}\times\big({\frac{1}{\hbar^2}\int_0^t dt_1 \mbox{Tr}_\varepsilon[\tilde{V}(t),[\tilde{V}(t_1),\tilde{\rho}(t)\otimes\rho_\varepsilon]]\big)
      e^{\frac{i}{\hbar}H_st}},
\label{QME}
\end{eqnarray}
where $\tilde{\rho}$ and $\tilde{V}$ are the density matrix of the
system and the interaction between the system and environments in
the interaction picture respectively, and ${\rm Tr}_\varepsilon$
indicates tracing over environment degrees of freedom with respect
to the thermal environment density matrix $\rho_\varepsilon$. In
obtaining Eq.\ (\ref{QME}), we have also used the fact that the
system-environment interaction is bilinear in their respective
positions (or displacements) so that the first-order term ${\rm
Tr}_\varepsilon( \tilde{V}(t)\rho_\varepsilon)=0$. We will derive
the non-Markovian master equations for our models using Eq.\
(\ref{QME}) without making any further approximation. For
simplicity, we assume the masses and resonance frequencies, and
coupling strengths to the environments are the same for the two
oscillators. That is $M_x=M_y=M$, $\Omega_x=\Omega_y=\Omega$ and
$\lambda_n^{(1)}=\lambda_n^{(2)}=\lambda_n$. We present the derived
non-Markovian master equations below.

{\bf Model A}: The master equation for two coupled oscillators, each
coupled to its own reservoir, of model A can be obtained as
% \begin{widetext}
  \begin{eqnarray}
      \dot{\rho}(t)&=&\frac{1}{i\hbar}[H_s+\frac{1}{4}M\tilde{\Omega}_1^2(t)(x-y)^2+\frac{1}{4}M\tilde{\Omega}_2^2(t)(x+y)^2,\rho]
\notag\\
%&&+\frac{1}{i\hbar}\frac{1}{4}M(\tilde{\Omega}_2^2(t)
%-\tilde{\Omega}_1^2(t))[xy+yx,\rho]\notag\\
      &&-\frac{i}{2\hbar}\gamma_1(t)[x-y,\{p_x-p_y,\rho\}]
\notag\\ &&-\frac{i}{2\hbar}\gamma_2(t)[x+y,\{p_x+p_y,\rho\}]\notag \\
      &&-\frac{1}{2}D_1(t)[x-y,[x-y,\rho]]
\notag\\ &&+\frac{1}{2}D_2(t)[x+y,[x+y,\rho]]\notag\\
      &&-\frac{1}{2\hbar}f_1(t)[x-y,[p_x-p_y,\rho]]
\notag\\&&+\frac{1}{2\hbar}f_2(t)[x+y,[p_x+p_y,\rho]],\label{master_eqA}
 \end{eqnarray}
% \end{widetext}
Here the time-dependent coefficients 
$\tilde{\Omega}_i^2(t)$ is called the frequency shift,
$\gamma_i(t)$ is the dissipation coefficient, and $D_i(t)$ and
$f_i(t)$ represent the diffusion coefficients. They can be written as 
 \begin{eqnarray}
      &&\tilde{\Omega}_i^2(t)=-\frac{2}{M} \int^t_0 dt' \cos(\Omega_i t')\eta(t'),\label{omega_tilde}\\
      &&\gamma_i(t)=\frac{1}{M \Omega_i} \int^t_0 dt' \sin(\Omega_i t')\eta(t'),\label{gamma}\\
      &&D_i(t)=\frac{1}{\hbar} \int^t_0 dt' \cos(\Omega_i t') \nu(t'),\label{D}\\
      &&f_i(t)=-\frac{1}{M \Omega_i}\int^t_0 dt' \sin(\Omega_i t') \nu(t'),\label{f}
 \end{eqnarray}
where $i=1,2$, and the frequencies $\Omega_1$ and $\Omega_2$ due to
the interaction $\lambda$ between the two oscillators are
\begin{eqnarray}
 \Omega_1&=&\sqrt{\Omega^2-\lambda/M},\\
 \Omega_2&=&\sqrt{\Omega^2+\lambda/M}.
 \end{eqnarray}
The two kernels $\eta(t')$ and
$\nu(t')$ appearing in Eqs.~(\ref{omega_tilde})--(\ref{f}) are,
respectively, the so-called dissipation and noise kernels and are
defined as
 \begin{eqnarray}
   \eta(t)&=&\frac{1}{2\hbar}\sum_n
   \lambda_n^2\langle[q_n(t),q_n(0)]\rangle \notag\\
&=&\int_0^\infty d\omega J(\omega)\sin(\omega t)\label{eta},\\
   \nu(t)&=&\frac{1}{2\hbar}\sum_n \lambda_n^2\langle\{q_n(t),q_n(0)\}\rangle\notag\\
&=&\int_0^\infty d\omega J(\omega)\cos(\omega
   t)(1+2N(\omega)),\label{nu}
 \end{eqnarray}
where
\begin{eqnarray}
  N(\omega)=\frac{1}{e^{\hbar\omega/{k_BT}}-1}
\end{eqnarray}
is the mean occupation number of the environmental oscillators and
\begin{equation}
J(\omega)=\sum_n
\frac{\lambda_n^2}{2m_n\omega_n}\delta(\omega-\omega_n)
\end{equation}
is the spectral density of the environments. Note that the
environment position operator $q_n(t)$ in Eqs.~(\ref{eta}) and
(\ref{nu}) should be $q_n^{(i)}(t)$ for different environments $i$.
We could in principle deal with this situation, but for simplicity
we assume that they have the same corresponding correlations even
though the two environments are independent of each other. It is
also worth noting that the frequency shift (\ref{omega_tilde}) and
dissipation coefficient (\ref{gamma}) depend only on the dissipation
kernel (\ref{eta}) while the diffusion coefficients, (\ref{D}) and
(\ref{f}), in the approximation of the perturbative expansion depend
only on the noise kernel (\ref{nu}), thus temperature dependent. We
can see from Eq.~(\ref{master_eqA}) that the term proportion to
$\gamma_i(t)$ is responsible for relaxation and the term
proportional to $D_i(t)$ is the main cause for decoherence. The
spectral density specifies the structure and properties of the
environment and thus determines the environmental influence on the
dynamics of the system of interest. In fact, the time evolution
behavior of the coefficients of the quantum master equation is
rather different for environments with different spectral content.

In principle, we could deal with any given form of the spectral
density. But as a particular example, we use the following form of
spectral density to specify the environments \cite{Leggett87, Hu}
 \begin{equation}
       J(\omega)=\frac{2}{\pi}M\gamma_0\omega\Big(\frac{\omega}{\Lambda}\Big)^{n-1}e^{-\omega^2/\Lambda^2},\label{spectral_density}
 \end{equation}
 where $\Lambda$ is the cutoff frequency, $\gamma_0$ is a constant
 characterizing the strength of the interaction with the environment,
 and $M$ is the system mass.
The environment is said to be ohmic if in the physical range of
frequencies
 ($\omega<\Lambda$) the spectral density is proportional to
 $\omega$. And it is said to be supra-ohmic if $J(\omega)$
 is proportional to $\omega^n$, $n>1$, or sub-ohmic if $n<1$.
For simplicity, in the following we focus on ohmic baths, i.e., the
case of $n=1$ in Eq.~(\ref{spectral_density}).

If we let
$\lambda=0$, then this model reduces to two non-interacting
oscillators, each coupled to its own reservoir. 
The master equation (\ref{master_eqA}) in this instance 
also reduces to the case of just putting two
sets of the quantum Brownian motion master equations
\cite{Paz_lecture} together:
 \begin{eqnarray}
      \dot{\rho}(t)&=&\frac{1}{i \hbar}[H_s+\frac{1}{2}M \tilde{\Omega}^2(t)(x^2+y^2),\rho]\notag \\
&&-\frac{i}{\hbar} \gamma(t) \big([x,\{p_x,\rho\}]+[y,\{p_y,\rho\}]\big)\notag\\
&&-D(t)\big([x,[x,\rho]]+[y,[y,\rho]]\big)\notag \\
&&-\frac{1}{\hbar}f(t)\big([x,[p_x,\rho]]+[y,[p_y,\rho]]\big),
\label{master_eqA0}
 \end{eqnarray}
where the time-dependent coefficients 
$\tilde{\Omega}^2(t)$, $\gamma(t)$, $D(t)$, and $f(t)$ 
are defined correspondingly as 
those in Eqs.(\ref{omega_tilde})-(\ref{f}) with $\lambda=0$.

{\bf Model B}: The master equation for two interacting oscillators,
coupled to a common reservoir, of model B can be derived and written as
% \begin{widetext}
 \begin{eqnarray}
\dot{\rho}(t)=&&\frac{1}{i \hbar}[H_s+\frac{1}{2}M \tilde{\Omega}_2^2(t)(x+y)^2,\rho]
\notag\\
%&&+\frac{1}{2i\hbar}M\tilde{\Omega}_2^2[xy+yx,\rho]\notag\\
      &&-\frac{i}{\hbar} \gamma_2(t) [x+y,\{p_x+p_y,\rho\}]
\notag\\ &&-D_2(t)[x+y,[x+y,\rho]]
 \notag\\    &&-\frac{1}{\hbar}f_2(t)[x+y,[p_x+p_y,\rho]].\label{master_eqB}
 \end{eqnarray}
% \end{widetext}
The time-dependent coefficients are the same as those in
Eqs.(\ref{omega_tilde})-(\ref{f}) with $i=2$ in model A. 
%Similarly, if we let the interaction strength $\lambda=0$,
%then this model is reduced to two non-interacting oscillators,
%coupled to a common reservoir. 
Note that compared with Eq.~(\ref{master_eqA}),
the mode $(x-y)$ is absent in Eq.~(\ref{master_eqB}).
This is a consequence of both the assumptions of
$M_x=M_y=M$, $\Omega_x=\Omega_y=\Omega$ and $\lambda_n^{(1)}=
\lambda_n^{(2)}=\lambda_n$ which we make to simplify the  
calculation as well as the nature of model B which is coupled to a
common bath with thus the variable $(x+y)$. 
This can also be inferred from an effective factor of two
difference in the corresponding coefficients of 
terms containing the $(x+y)$ variable between 
Eq.~(\ref{master_eqA}) and Eq.~(\ref{master_eqB}). 
In the derivation of the master equation in model B, some sort of
additions make the coefficients of $(x+y)$ mode twice larger and some
sort of cancellations make the $(x-y)$ mode absent in
Eq.~(\ref{master_eqB}).    
We note that despite being derived
perturbatively, the master equations for the two models seem to be
very similar to their exact counterparts which are also
time-convolutionless with time-dependent coefficients \cite{Hu,Chou07}.

%%%%%%%%%%%%%%%%%%%%%%%%%%%%%%%%%%%%%%%%%%%%%%%%%%%%%%%%%%%%%%%%%%
\section{Renormalization and time-dependent coefficients}\label{sec42}
%%%%%%%%%%%%%%%%%%%%%%%%%%%%%%%%%%%%%%%%%%%%%%%%%%%%%%%%%%%%%%%%%%%

We note that due to
the interaction with the environment, the frequency shift term,
 $\tilde{\Omega}_i^2(t)$ of Eq.\ (\ref{omega_tilde}),
%,  is proportional to the product $\gamma_0 \Lambda$ and thus
diverges as the cutoff frequency $\Lambda\rightarrow\infty$ and thus
     is not physical.
Thus a regularization procedure is needed for the frequency renormalization.
There are two different views on this renormalization
     \cite{Ryder96,Mandl93}.  One view is starting
     from the original Hamiltonian, the frequency can be made finite, by
     a renormalization of frequency, from its {\it bare} to its
     {\it physical} value.
Thus by combining terms which involve the frequency in the master
equation, the physical frequency which is the quantity that can be
measured in the laboratory is defined as
     \begin{align}
      \Omega_p^2(t)=\Omega^2+\tilde{\Omega}^2(t),
     \end{align}
     where $\Omega$ is the bare frequency in the Hamiltonian.
In this case, the physical frequency is taken to be finite and
the bare frequency is taken
     to be infinite as $\Lambda\to \infty$ in order to cancel the divergent
     contribution from  $\tilde{\Omega}^2(t)$.
Thus the bare frequency has no direct physical significance.
Although it may not be exactly the same,
this view of renormalization has an analogy
     in solid state physics where, for example, electrons are attributed
     an {\it effective mass} to take into account their interaction with
     the lattice or/and other electrons.

An alternative view of renormalization is to regard the frequency
$\Omega$ in the original Hamiltonian as a finite renormalized
frequency $\Omega_r$. The fact that this Hamiltonian does not give finite
frequency then requires that extra terms be added to the Hamiltonian
to cancel the divergence. These terms are called
counter-terms  \cite{Ryder96,Mandl93,Leggett83,Weiss99}.
This view of renormalization
is in fact more commonly adopted in high energy physics or
quantum field theory.
In the context of a quantum Brownian motion model,
the total Hamiltonian in this case
can be written as
\begin{eqnarray}
      H&=&{\frac{p^2}{2M}+\frac{1}{2}M\Omega_r^2x^2}
      +\sum_n\Big(\frac{p_n^2}{2m_n}\notag\\
&&+\frac{1}{2}m_n\omega_n^2q_n^2
         +\lambda_n q_n x\Big)
 +\sum_n \frac{1}{2}\frac{\lambda_n^2}{m_n\omega_n^2}x^2
\label{H_ct1} \\
&=&{\frac{p^2}{2M}+\frac{1}{2}M\Omega_r^2x^2}\notag\\
      &&+\sum_n\Big[\frac{p_n^2}{2m_n}
+\frac{1}{2}m_n\omega_n^2\big(q_n+\frac{\lambda_n}{m_n\omega_n^2}x\big)^2\Big].
\label{H_ct2}
     \end{eqnarray}
The last term in Eq.~(\ref{H_ct1})
can be viewed as a frequency counter-term \cite{Leggett83,Weiss99} with a
frequency defined as
     \begin{align}
      \Omega_c^2=\frac{1}{M}\sum_n\frac{\lambda_n^2}{m_n\omega_n^2}
=2\int_0^\infty d\omega \frac{J(\omega)}{\omega}.
     \end{align}
%Equation (\ref{H_ct1}) can be written as a more concise form as
%     \begin{eqnarray}
%      H&=&{\frac{p^2}{2M}+\frac{1}{2}M\Omega^2x^2}\notag\\
%      &&+\sum_n\Big[\frac{p_n^2}{2m_n}
%+\frac{1}{2}m_n\omega_n^2\big(q_n+\frac{\lambda_n}{m_n\omega_n^2}x\big)^2\Big]
%\label{H_ct2}
%     \end{eqnarray}
We can find that at large times the frequency shift 
$\tilde{\Omega}^2(t)$ is negative
and equals to $-\Omega_c^2$. The added counter-term is to cancel the
frequency shift at long times and ensure that the system can not
lower its potential energy below the original (renormalized) value.
The physical frequency in this case is
$\Omega_p^2=\Omega_r^2+\Omega_c^2+\tilde{\Omega}^2(t)$ and equals
     to  $\Omega_r^2$ at long times.
We will take this view of renormalization, so the original frequency
$\Omega$ in the time-dependent coefficients of the master equations,
Eqs.~(\ref{master_eqA}) and (\ref{master_eqB}), should be replaced
by $\Omega\rightarrow\Omega_r$.

%%%%%%%%%%%%%%%%%%%%%%%%%%%%%%%%%%%%%%%%%%%%%%%%%%%%%%%%%%
\section{\label{sec4} Logarithmic negativity and two-mode Gaussian states}
%%%%%%%%%%%%%%%%%%%%%%%%%%%%%%%%%%%%%%%%%%%%%%%%%%%%%%%%%%%%%%%%%%

The purpose of this paper is to focus on the entanglement dynamics
for the reduced density matrix of the two oscillators in our models.
%Therefore a computable measure of entanglement for
%infinite-dimensional bipartite system states is needed. 
The derived
master equations are generally partial differential equations with
time-dependent coefficients. Consequently, computing the time
evolution solution for the density matrix operator explicitly and
then using it to calculate directly the dynamics of entanglement
%negativity through Eq.~(\ref{logE}) 
are still considered difficult.
%For continuous variable systems,
%the problems become complicated as we need to solve the time evolution
%of the partial differential equation of the density matrix in order to
%compute the entanglement dynamics.
But the problem becomes much more tractable if we restrict the states
to be Gaussian states.
Since the couplings in our models are all bilinear in their respective
positions (displacements) and the effects of the environments in the
master equations have operator structure no more than quadratic in
the momenta or/and  positions of the two oscillators,
an initial Gaussian state would remain
Gaussian in its subsequent time evolution. 
So, for simplicity, we will consider in the following
Gaussian initial states for the two oscillators.
Any Gaussian state can be completely
characterized by its corresponding covariance matrix.  We will see
below that the time evolution of the covariance matrix is easier to
calculate than that of the density matrix.

A set of Gaussian states is the set of states
with Gaussian characteristic functions and quasi-probability
distributions of the Wigner function \cite{Adesso}.
The Wigner quasi-probability distribution function is defined in
terms of density matrix $\rho(t)$ as:
 \begin{equation}
       W(\textbf{\emph{q}},\textbf{\emph{p}},t)\equiv
\Big(\frac{1}{2\pi\hbar}\Big)^n\int_{-\infty}^\infty
       d\bm{\xi}
       \langle \textbf{\emph{q}}-\frac{\bm{\xi}}{2}\vert{\rho}(t)\vert \textbf{\emph{q}}+\frac{\bm{\xi}}{2}
       \rangle\exp(\frac{i\textbf{\emph{p}}\cdot\bm{\xi}}{\hbar}).
\label{wigner_function_n}
 \end{equation}
So a zero-mean Gaussian state is described, for example, by the Wigner
function as
\begin{equation}
W(\mathbf{X})=\frac{1}{4\pi^n\sqrt{\mbox{det}\mathbf{V}}} \exp{(-\frac{1}{2} \mathbf{X} \mathbf{V}^{-1}
\mathbf{X}^T)},
\label{Gaussian_wf}
\end{equation}
where $\mathbf{V}$ is the covariance matrix and $\mathbf{X}$
represents the vector $(x_1, p_1, x_2, p_2,\cdots, x_n, p_n)$. The
zero mean denotes $\langle X_i\rangle=0$, and this can be changed at
will using local unitary displacement operators. So we can set
$\langle X_i\rangle=0$ without loss of generality.

The matrix elements of the covariance matrix $\mathbf{V}$ are defined as
  \begin{eqnarray}
       V_{i,j}&\equiv&\langle\{\Delta\hat{X}_i,\Delta\hat{X}_j\}\rangle
       =\mbox{Tr}\big(\{\Delta\hat{X}_i,\Delta\hat{X}_j\}\hat{\rho}\big)\notag\\
       &=&\int d^4X\mbox{ }\Delta X_i\Delta X_j W(X),\label{covariance_matrix}
  \end{eqnarray}
  where $\{\Delta\hat{X}_i,\Delta\hat{X}_j\}=(\Delta\hat{X}_i\Delta\hat{X}_j+\Delta\hat{X}_j\Delta\hat{X}_i)/2$,
  $\Delta\hat{X}_i=\hat{X}_i-\langle\hat{X}_i\rangle$ and $\Delta X_i=X_i-\langle X_i\rangle$.
The average of the operator $\hat{X}_i$,
$\langle\hat{X}_i\rangle$, means
  $\mbox{Tr}\big(\hat{X}_i\hat{\rho}\big)$, and $\langle X_i\rangle$
  denotes an
  average of a variable $X_i$ with respect to the Wigner function
  distribution $W(\mathbf{X})$, so $\langle\hat{X}_i\rangle$ equals to
  $\langle X_i\rangle$.
With this definition, we could transfer the problem of solving a
time-dependent partial
differential equation of the density matrix into a problem of solving
first-order in time, coupled linear ordinary differential equations of the
covariance matrix elements.  This could be done by first
transferring the master equations to the Fokker-Planck equation for
the Wigner function, and then finding the coupled differential
evolution equation for the covariance matrix elements using Eq.\
(\ref{covariance_matrix}) and
the Fokker-Planck equation (see Sec.\ \ref{sec41}).

We will use
the logarithmic negativity to quantify the degrees of entanglement
of the infinite-dimensional bipartite system states of the two oscillators.
The logarithmic negativity of a bipartite system consisting of two
subsystems A and B is \cite{Vidal}
  \begin{equation}
   E_N(\rho)\equiv \log_2{\parallel \rho^{T_B}\parallel_1},
\label{logE}
  \end{equation}
where $\rho^{T_B}$ means partial transpose of a (mixed)
state density matrix operator $\rho$
with respect to subsystem $B$. That is to say,
  $\langle i_A,j_B\vert\rho^{T_B}\vert k_A,l_B\rangle\equiv\langle i_A,l_B\vert\rho\vert k_A,j_B\rangle$
  for any arbitrary orthonormal product basis which is belonged to the tensor product of Hilbert space of combinative
system $A$ and $B$.
The operation $\parallel . \parallel_1$ denotes the trace norm and
the trace norm of any Hermitian operator
  $H$ is defined as $\parallel H\parallel_1\equiv \mbox{Tr}\vert
  H\vert\equiv \mbox{Tr}\sqrt{H^\dag H}$.

Despite not being convex, 
the logarithmic negativity is a full entanglement monotone under
local operations and classical communication \cite{Plenio05}
and constitutes to an upper bound to the {\it distillable
  entanglement} \cite{Vidal}.  
For the particular case of two-mode Gaussian states,
the logarithmic negativity could actually provide an appropriate
quantification of quantum entanglement.
Vidal and Werner \cite{Vidal}
demonstrated that logarithmic
negativity is computable for general Gaussian states.
For two-mode Gaussian states, it can be furthermore shown that 
the logarithmic negativity can be represented as \cite{Adesso} 
%is a simple function of the smallestThe smaller the value of $V_s$ 
%partially transposed sympletic eigenvalue $V_s$: 
\begin{equation}
       E_N(\rho)
       =\mbox{max}\big(0,-\log_2 2 V_s \big).
\label{etanglement_V}
  \end{equation}
where $V_s$ is the smallest sympletic eigenvalue of the partially
transposed covariance matrix of the two-mode Gaussian states. 
Equation (\ref{etanglement_V}) is a simple decreasing function of $V_s$ 
which quantifies the degree of violation of the necessary and sufficient
separability criterion of the positivity of partial transpose
\cite{Simon00,Adesso}.  
For $V_s\geq 1/2$ the state is separable, otherwise it is entangled.  
So the smallest partially transposed sympletic eigenvalue $V_s$ alone
completely qualifies and quantifies the quantum entanglement of a
two-mode Gaussian state  \cite{Adesso}. 
That is, the smaller the value of $V_s$, the more entangled the
corresponding two-mode Gaussian state.   
As a result, the
logarithmic negativity may be regarded as a suitable entanglement
quantification indicator for two-mode Gaussian states. 

The partially transposed sympletic eigenvalues 
$V_i$ are the symplectic eigenvalues of $\mathbf{V}^{T_B}$, and
$\mathbf{V}^{T_B}$ can be written down in a compact form \cite{Plenio04}
  \begin{equation}
       \mathbf{V}^{T_B}=P\mathbf{V}P\label{partial_tranpose_V},
  \end{equation}
  where
  \begin{equation}
       P=
       \left(
         \begin{array}{cc}
           1 & 0 \\
           0 & 1 \\
         \end{array}
       \right)\oplus
       \left(
         \begin{array}{cc}
           1 & 0 \\
           0 & -1 \\
         \end{array}
       \right),
  \end{equation}
and $A\oplus B$ means that the block-diagonal matrix with the matrices $A$
and $B$ as diagonal entries.
The symplectic eigenvalues are the positive square roots of
  the standard eigenvalues of the $-\sigma \mathbf{V}^{T_B}\sigma
  \mathbf{V}^{T_B}$ or the absolute value of
the eigenvalues of $i\sigma \mathbf{V}^{T_B}$.
Here $\sigma$ is called the symplectic matrix from the commutation relations
  $[\hat{x}_i,\hat{x}_j]=i\hbar\sigma_{i,j}$ which is given by
  \begin{equation}
      \sigma=
      \left(
        \begin{array}{cc}
          J & 0 \\
          0 & J \\
        \end{array}
      \right),\mbox{ }\mbox{ }\mbox{and}\mbox{ }\mbox{ }\mbox{ }
      J=
      \left(
        \begin{array}{cc}
          0 & 1 \\
          -1 & 0 \\
        \end{array}
      \right).
  \end{equation}

The logarithmic negativity in the form of Eq.\ (\ref{etanglement_V})
is much more easier to compute than that defined in Eq.\
(\ref{logE}). 
%In the following, we will use the logarithmic negativity 
%to quantify the quantum entanglement of two-mode Gaussian states. 

%%%%%%%%%%%%%%%%%%%%%%%%%%%%%%%%%%%%%%%%%%%%%%%%%%%%%%%%%%%%%%%%%%%%%
%\section{\label{sec3} Covariance
%  matrix and two-mode squeezed vacuum state}
%%%%%%%%%%%%%%%%%%%%%%%%%%%%%%%%%%%%%%%%%%%%%%%%%%%%%%%%%%%%%%%%%%%%

%For our purpose, we will first derive the non-Markovian quantum
%masters equation for the reduced density matrix operator of the two
%oscillators in thermal environments for our four models.
%The derived
%master equations are generally a partial differential equations with
%time-dependent coefficients. Consequently, computing the time
%evolution solution for the density matrix operator explicitly and
%then using it to calculate the dynamics of the logarithmic
%negativity using Eq.\ (\ref{logE}) is considered difficult.

One subclass of two-mode Gaussian states is the so-called two-mode squeezed
vacuum states. The position and momentum wave functions for the
two-mode squeezed 
vacuum state with a squeezing parameter $r$ are \cite{Braunstein05}
%\begin{eqnarray}
%\psi(x,y)&=&\sqrt{\frac{2}{\pi}}\exp[-e^{-2r}(x+y)^2/2
%\notag\\
%&&-e^{2r}(x_1-x_2)^2/2],
%  \label{eq:wfx}\\
%\bar{\psi}(p_1,p_2)&=&\sqrt{\frac{2}{\pi}}\exp[-e^{-2r}(p_1-p_2)^2/2
%\notag\\
%&&-e^{2r}(p_1+p_2)^2/2].
%  \label{eq:wfp}
%\end{eqnarray}
\begin{eqnarray}
\psi(x,y)&=&\sqrt{\frac{2}{\pi}}\exp[-e^{-2r}(x+y)^2/2
\notag\\
&&-e^{2r}(x-y)^2/2],
  \label{eq:wfx}\\
\bar{\psi}(p_x,p_y)&=&\sqrt{\frac{2}{\pi}}\exp[-e^{-2r}(p_x-p_y)^2/2
\notag\\
&&-e^{2r}(p_x+p_y)^2/2].
  \label{eq:wfp}
\end{eqnarray}
They approach $C\delta(x-y)$ and $C\delta(p_x+p_y)$ ,
respectively, in the limit of infinitely squeezing $r\to\infty$,
where $C$ is some constant.
The corresponding Wigner function of the two-mode squeezed
vacuum state is then \cite{Braunstein05}
%\begin{eqnarray}
%W(\mathbf{X})&=&\frac{4}{\pi^2}\exp\{-e^{-2r}[(x_1+x_2)^2+(p_1-p_2)^2]
%\notag\\
%&&-e^{2r}(x_1-x_2)^2+(p_1+p_2)^2]\}.
%\label{eq:Wigner}
%\end{eqnarray}
\begin{eqnarray}
W(\mathbf{X})&=&\frac{4}{\pi^2}\exp\{-e^{-2r}[(x+y)^2+(p_x-p_y)^2]
\notag\\
&&-e^{2r}(x-y)^2+(p_x+p_y)^2]\}.
\label{eq:Wigner}
\end{eqnarray}
In the limit of infinitely squeezing $r\to\infty$,
this Wigner function approaches  $C\delta(x-y)\delta(p_x+p_y)$,
corresponding to the original (perfectly correlated and maximally
entangled) EPR state.
While at $r=0$, the two-mode state corresponds to a separable
(disentangled) state.
The two-mode squeezed
vacuum states are routinely generated in quantum optics laboratories
and have been used in most implementations of continuous variable
quantum information protocols \cite{Furusawa98,Laurat,Braunstein05}.
It has also been proposed recently that a two-mode squeezed state could
be generated for two nanomechanical oscillators that act as the two opposite
sections, suspended from the substrate, of a dc-SQUID
(superconducting quantum interference device) loop \cite{Xue07}.

The two-mode squeezed
vacuum states, from Eqs.~(\ref{Gaussian_wf}) and (\ref{eq:Wigner}), can
be completely characterized by the following
covariance matrix \cite{Simon00,Duan}:
  \begin{equation}
       \mathbf{V}\equiv
       \left(
        \begin{array}{cccc}
          a & 0 & -c & 0 \\
          0 & a & 0 & c \\
          -c & 0 & a & 0 \\
          0 & c & 0 & a \\
        \end{array}
       \right),
  \end{equation}
where $a$ and $c$ are
  \begin{equation}
       a={\cosh(2r)}/{2},\mbox{ }\mbox{ }\mbox{ }\mbox{ }\mbox{
       }\mbox{ } c={\sinh(2r)}/{2}.
  \end{equation}
%Here $r$ is the squeezing parameter, and $r$ equals to zero
%  corresponds to a separable (disentangled) state.
For simplicity, we will use the two-mode squeezed vacuum states as the initial
states of the two quantum oscillators in our models throughout the paper.

%%%%%%%%%%%%%%%%%%%%%%%%%%%%%%%%%%%%%%%%%%%%%%%%%%%%%%%%%%%%%%%%
%\subsection{Fokker Planck equation}
\section{\label{sec41} Evolution equations of the Covariance Matrix Elements}
%%%%%%%%%%%%%%%%%%%%%%%%%%%%%%%%%%%%%%%%%%%%%%%%%%%%%%%%%%%%%%%%%

By the definition of Wigner function in Eq.~(\ref{wigner_function_n}),
the corresponding function for $\dot{\rho}(t)$ is
\begin{equation}
       \dot{W}(\textbf{\emph{q}},\textbf{\emph{p}},t)=\small{\Big(\frac{1}{2\pi\hbar}\Big)^n\int_{-\infty}^\infty d\bm{\xi}
       \langle \textbf{\emph{q}}-\frac{\bm{\xi}}{2}\vert\dot{\rho}(t)\vert \textbf{\emph{q}}+\frac{\bm{\xi}}{2}
       \rangle\exp(\frac{i\textbf{\emph{p}}\cdot\bm{\xi}}{\hbar})}.
\label{Wdot}
 \end{equation}
Using Eqs.\ (\ref{wigner_function_n}) and (\ref{Wdot}), and after
straightforward but somehow tedious calculations, we can obtain the
Fokker-Planck equations of the Wigner function corresponding to the
master equations (\ref{master_eqA}) and (\ref{master_eqB}). From the
Fokker-Planck equation of the Wigner function and Eq.\
(\ref{covariance_matrix}), we can obtain coupled first-order
ordinary differential equations with time-dependent coefficients for
all elements of the covariance matrix. Due to the symmetrical
property of the covariance matrix, i.e. $V_{12}=V_{21}$,
$V_{24}=V_{42}$ etc., we only need ten essential components of the
covariance matrix in the bipartite system here instead of sixteen
components. 
%The covariance matrix representation,
%for example for Eq.\ (\ref{F_P_eqA}) of Model A with
%renormalized frequency, is
For example, we obtain for model B with the renormalized frequency
$\Omega\to \Omega_r$:
\begin{eqnarray}
          \dot{V}_{11}&=&2V_{12},\notag\\
          \dot{V}_{12}&=&-(\Omega_r^2+\Omega_c^2+\tilde{\Omega}_2^2(t))V_{11}-2\gamma_2(t)V_{12}\notag\\
         &&-(\Omega_c^2+\tilde{\Omega}_2^2(t)+\lambda/M)V_{13}
-2\gamma_2(t)V_{14}+V_{22}-\hbar f_2(t),\notag\\
          \dot{V}_{13}&=&V_{14}+V_{23},\notag\\
          \dot{V}_{14}&=&-(\Omega_c^2+\tilde{\Omega}_2^2(t)+\lambda/M)V_{11}-2\gamma_2(t)V_{12}\notag\\
&&-(\Omega_r^2+\Omega_c^2+\tilde{\Omega}^2_2(t))V_{13}
          -2\gamma_2(t)V_{14}+V_{24}-\hbar f_2(t),\notag\\
          \dot{V}_{22}&=&-2(\Omega_r^2+\Omega_c^2+\tilde{\Omega}_2^2(t))V_{12}-4\gamma_2(t)V_{22}\notag\\
          &&-2(\Omega_c^2+\tilde{\Omega}_2^2(t)+\lambda/M)V_{23}-4\gamma_2(t)V_{24}+2\hbar^2D_2(t),\notag\\
          \dot{V}_{23}&=&-(\Omega_r^2+\Omega_c^2+\tilde{\Omega}_2^2(t))V_{13}-2\gamma_2(t)V_{23}+V_{24}\notag\\
         &&-(\Omega_c^2-\tilde{\Omega}_2^2+\lambda)V_{33}-2\gamma_2V_{34}-\hbar f_2(t),\notag\\
          \dot{V}_{24}&=&-(\Omega_c^2+\tilde{\Omega}_2^2(t)+\lambda/M)V_{12}-(\Omega_r^2+\Omega_c^2+\tilde{\Omega}_2^2(t))V_{14}\notag\\
          &&-2\gamma_2(t)V_{22}
-(\Omega_r^2+\Omega_c^2+\tilde{\Omega}_2^2(t))V_{23}-4\gamma_2(t)V_{24}\notag\\
         && -(\Omega_c^2+\tilde{\Omega}_2^2(t)+\lambda/M)V_{34}
-2\gamma_2(t)V_{44}+2\hbar^2D_2(t),\notag\\
          \dot{V}_{33}&=&2V_{34},\notag\\
          \dot{V}_{34}&=&-(\Omega_c^2+\tilde{\Omega}_2^2(t))V_{13}-2\gamma_2(t)V_{23}\notag\\
&&-(\Omega_r^2+\Omega_c^2+
          \tilde{\Omega}_2^2(t))V_{33}-2\gamma_2(t)V_{34}+V_{44}-\hbar f_2(t),\notag\\
          \dot{V}_{44}&=&-2(\Omega_c^2+\tilde{\Omega}_2^2(t)+\lambda/M)V_{14}-4\gamma_2(t)V_{24}\notag\\
         &&-2(\Omega_r^2+\Omega_c^2+\tilde{\Omega}_2^2(t))V_{34}-4\gamma_2(t)V_{44}+2\hbar^2D_2(t).\notag
\end{eqnarray}
Similar calculations can be performed for model A. 
The solutions of the coupled first-order ordinary
differential equations in time are much easier to calculate than the
partial differential equations of the Fokker-Planck equations or the
quantum master equations. Solving for the time evolution of the
covariance matrix elements, we can then obtain the entanglement
dynamics through the computation of the logarithmic negativity using
Eq.\ ({\ref{etanglement_V}). 
%Similar calculations can be performed
%for other master equations in the models that we consider.

%%%%%%%%%%%%%%%%%%%%%%%%%%%%%%%%%%%%%%%%%%%%%%%%%%%%%%%%%%%%%%%%%
\section{\label{sec5}Results and discussions}
%%%%%%%%%%%%%%%%%%%%%%%%%%%%%%%%%%%%%%%%%%%%%%%%%%%%%%%%%%%%%%%%%%
We first report our numerical results of the non-Markovian
entanglement (logarithmic negativity)
  dynamics for our two models in two cases: (1)  $\gamma_0=0$ (isolated),
%(2) $\gamma_0=6\times10^{-4}\Omega_r$ (week coupling) and 
(2) $\gamma_0=6\times10^{-2}\Omega_r$,  where
$\gamma_0$ is a constant in the spectral density
(\ref{spectral_density}) and is related to the coupling strength to
the environments and $\Omega_r$ is the renormalized frequency of the
subsystems. In all the plots presented below, the parameters used
are as follows. The environment temperature is at $k_B
T=10\hbar\Omega_r$, the cutoff
  frequency is $\Lambda=2000\Omega_r$ and
the interaction between two subsystems $\lambda$ is in units of $M\Omega_r^2$.
Finally, we investigate whether the
entanglement survival condition under RWA-Markovian approximation in
Ref. \cite{Jakub} is still valid in the non-Markovian case.

\subsection{Isolated system ($\gamma_0=0$)}
\begin{figure}
   \includegraphics[width=8.5cm]{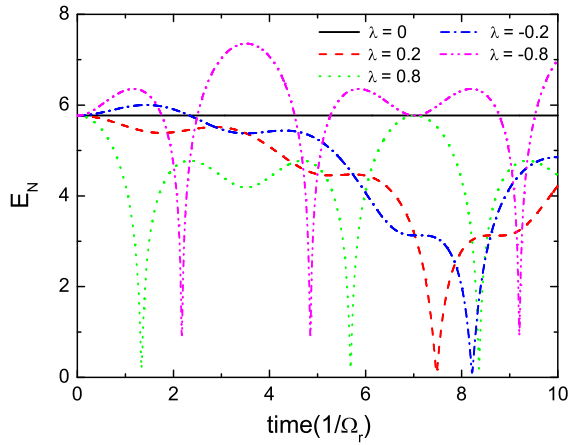}
       \caption{(Color online). Time evolution of the logarithmic negativity of the two subsystems isolated from external environments
        ($\gamma_0=0$) for the case of an initial two-mode squeezed vacuum state with a
        squeezing parameter $r=2$. The solid line stands for $\lambda=0$, dashed for $\lambda=0.2$,
        dotted for $\lambda=0.8$, dash-dotted for $\lambda=-0.2$ and dash-dot-dotted for $\lambda=-0.8$.
        In all the plots presented below, the parameters used are as follows.
        The environment temperature is at $k_B T=10\hbar\Omega_r$, the cutoff
        frequency is $\Lambda=2000\Omega_r$ and
        the interaction between two subsystems $\lambda$ is in units of $M\Omega_r^2$.
        }\label{fig_iso_squeez2}
  \end{figure}
\begin{figure}
 \includegraphics[width=8.5cm]{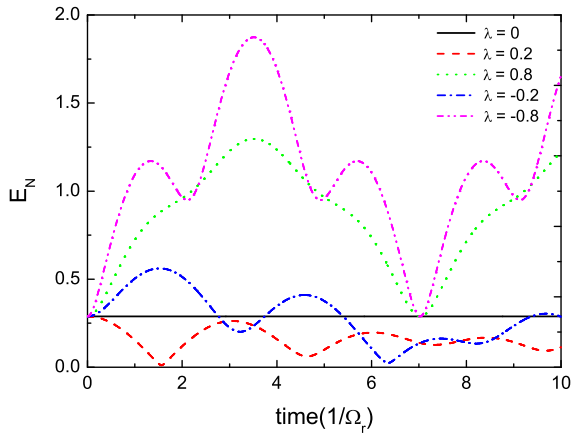}
       \caption{(Color online). Time evolution of the logarithmic negativity for an initial two-mode squeezed vacuum state with a
        squeezing parameter $r=0.1$. Other conditions and plot caption are the same as in
        Fig.\ \ref{fig_iso_squeez2}.
        }\label{fig_iso_squeez01}
\end{figure}Similar calculations can be performed for 
model A. 

  \begin{figure}
   \includegraphics[width=8.5cm]{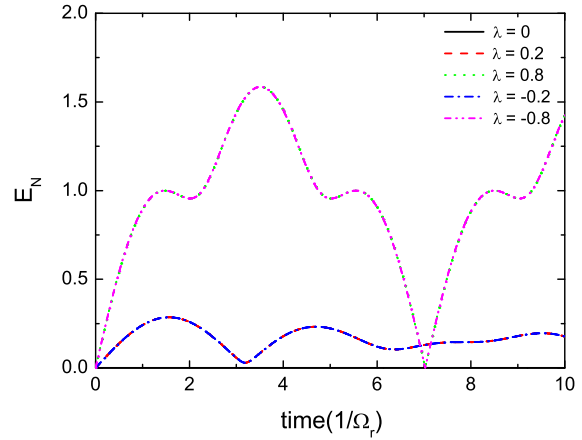}
       \caption{(Color online). Time evolution of the logarithmic negativity for an initial two-mode squeezed vacuum state with a
        squeezing parameter $r=0$. Other conditions and plot caption
are the same as in
        Fig.\ \ref{fig_iso_squeez2}.
        }\label{fig_iso_squeez0}
  \end{figure}

We plot the dynamics of the logarithmic negativity of the two models
when they are isolated from the external environments in
Figs.~\ref{fig_iso_squeez2}--\ref{fig_iso_squeez0}. 
Similar to that of an interacting discrete system of two qubits, the
dynamics of entanglement of the two oscillators depends strongly on
the initial states and on the interacting strength between them.  
When there is
no interaction between the two subsystems ($\lambda=0$) isolated from
the environments, the time evolution of the
logarithmic negativity maintains constant in
Figs.~\ref{fig_iso_squeez2}--\ref{fig_iso_squeez0} as it should.
But the dynamics of the logarithmic negativity varies
quasi-periodically for two interacting subsystems isolated from the
external environments and the smaller the value of the interaction strength
  $|\lambda|$ between the two oscillators is, the longer the quasi-period
  of the logarithmic negativity is. This can be seen from the plots of the
  entanglement dynamics in a longer time scale.

From Fig.~\ref{fig_iso_squeez0}, we see that the entanglement of the
  two subsystems can be generated from an
  initially separable state ($r=0$) through their
  mutual interaction, and the larger the interaction strength, the
  larger the generation of the entanglement.
We can also see from  Fig.~\ref{fig_iso_squeez0} that the
entanglement dynamics for an initially separable
state ($r=0$) seems to be
symmetric with respect to the change of $\lambda xy
  \leftrightarrow -\lambda xy$. 
While this is not the case for  $r\neq 0$ ( see 
Figs.~\ref{fig_iso_squeez2} and \ref{fig_iso_squeez01}). 
This may be due to the fact that for $r=0$ the
initial wave function, Eq.~(\ref{eq:wfx}), or
the Wigner function, Eq.~(\ref{eq:Wigner}), is symmetric under the
  change of
$x \leftrightarrow -x$, $y \leftrightarrow -y$ and
$xy \leftrightarrow -xy$.
While for $r\neq 0$, this symmetry is broken and the
initial wave function, Eq.~(\ref{eq:wfx}), or
the Wigner function, Eq.~(\ref{eq:Wigner}),
possesses the preferred entanglement in
the relative position variable $(x-y)$ as compared to the variable $(x+y)$.
%As a result, the $\lamda<0$
So an attractive interaction ($\lambda<0$) seems to enhance this
entanglement in $(x-y)$. This can be seen from
Figs.~\ref{fig_iso_squeez2}--\ref{fig_iso_squeez0} that for a fixed
value of the squeezing parameter $r$, if the
  interaction strength is attractive ($\lambda<0$) then the
entanglement grows initially with time.
On the other hand, the entanglement
  decreases with time initially if the
  interaction strength is positive and smaller than a
  critical value, i.e. $0 < \lambda < \lambda_c$.
For example, in Fig.~\ref{fig_iso_squeez01} the initial entanglement
  grows with time for $\lambda=0.8$
while it decreases with time for $\lambda=0.2$.
Similarly, we may say that
for a fixed positive value of $\lambda>0$, there exists a
critical initial squeezing parameter above which the entanglement
  decreases with time initially.
Figure \ref{critical_r} shows the critical value
  line that separates these two situations in the
positive interaction strength $\lambda$ versus initial squeezing parameter
$r$ phase diagram.
\begin{figure}
   \includegraphics[width=8cm]{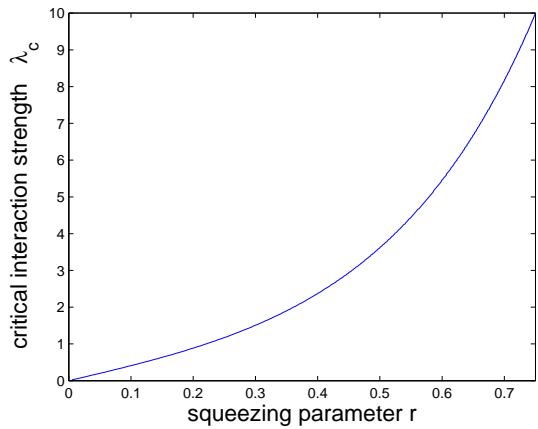}
   \caption{The positive interaction strength versus
initial squeezing parameter phase diagram. The regime above (below) the
critical line curve corresponds to the situation that the entanglement
increases (decreases) with time initially.}\label{critical_r}
\end{figure}

The general trend is that when the entanglement grows with time
  initially, the entanglement is enhanced to reach maximum values at
  later times; while if the entanglement decreases with time initially,
  the initial value of the entanglement is usually the maximum value.
Thus, for two oscillators initially in a two-mode squeezed vacuum state
  with a squeezing parameter $r$,
an attractive interaction ($\lambda<0$) 
between the two oscillators is able to enhance their
entanglement at later times. 
An repulsive interaction ($\lambda>0$) can enhances
the entanglement at later times if $\lambda>\lambda_c$, but the
entanglement is no longer increased if $0 <\lambda < \lambda_c$.
For the same value of interaction strength between the two oscillators,
the attractive interaction seems always to be better than the repulsive
  interaction as far as the maximum value of entanglement that can be
  reached at a later time is concerned.

%%%%%%%%%%%%%%%%%%%%%%%%%%%%%%%%%%%%%%%%%%%%%%%%%%%%%%%%%%%%%%%
\subsection{Coupled to environments ($\gamma_0=6\times10^{-2}\Omega_r$)}
%%%%%%%%%%%%%%%%%%%%%%%%%%%%%%%%%%%%%%%%%%%%%%%%%%%%%%%%%%%%%%%

\begin{figure}
  \includegraphics[width=8.5cm]{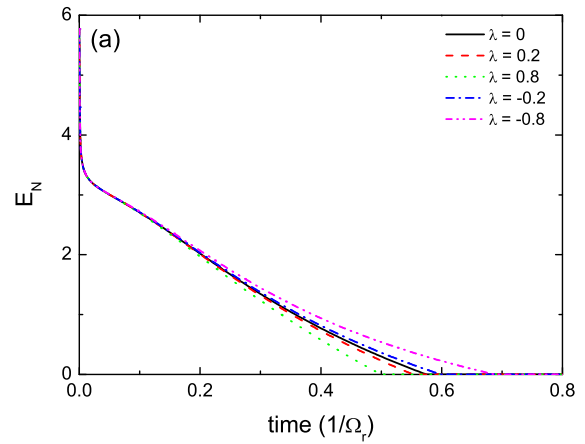}
  \includegraphics[width=8.5cm]{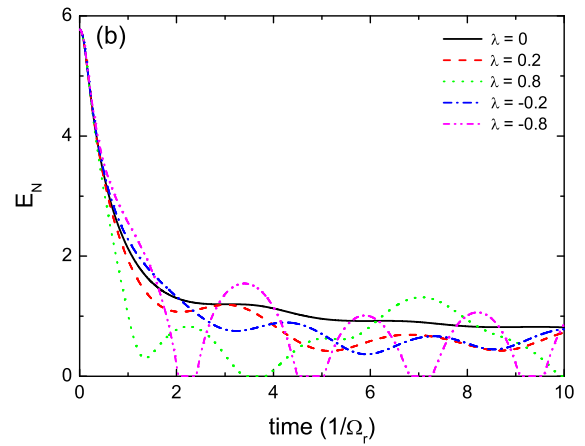}
  \caption{(Color online). Time evolutions of the logarithmic negativity of the two
    subsystems coupled more strongly to the environments ($\gamma_0=6\times10^{-2}\Omega_r$)
        for an initial two-mode squeezed vacuum state with a squeezing parameter $r=2$.
        The subplot (a) is for Model A and (b) is for Model B.
   The solid line is for
   $\lambda=0$, dashed for $\lambda=0.2$, dotted for $\lambda=0.8$, dash-dotted for $\lambda=-0.2$ and dash-dot-dotted for
   $\lambda=-0.8$.}\label{fig_ohmic_LSab1}\label{fig_ohmic_LScd1}
  \end{figure}
  \begin{figure}
   \includegraphics[width=8.8cm]{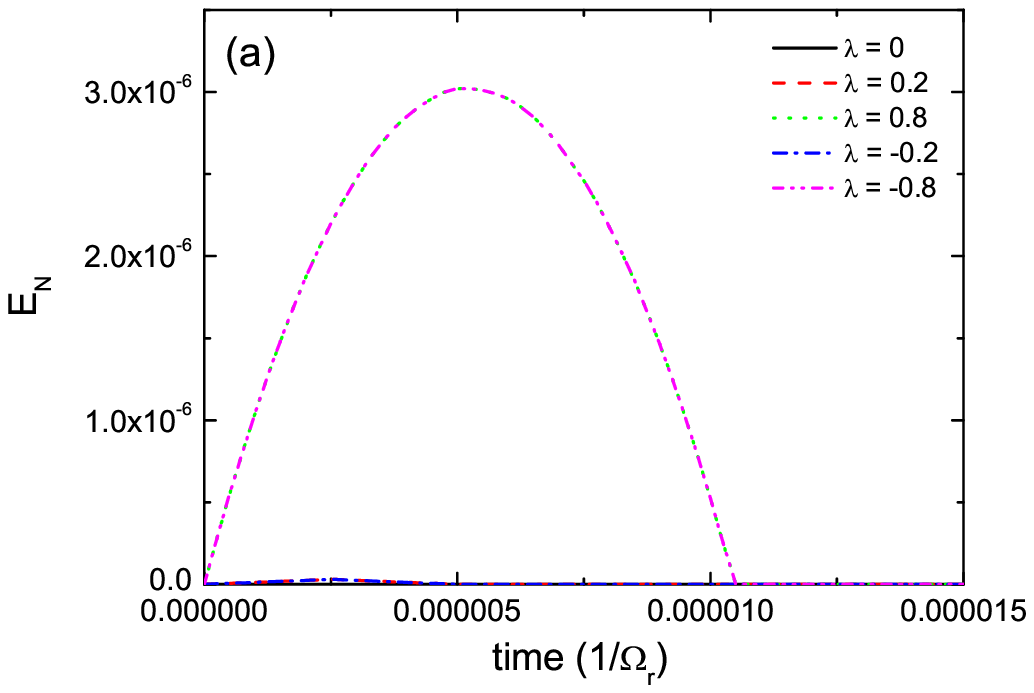}
   \includegraphics[width=8.5cm]{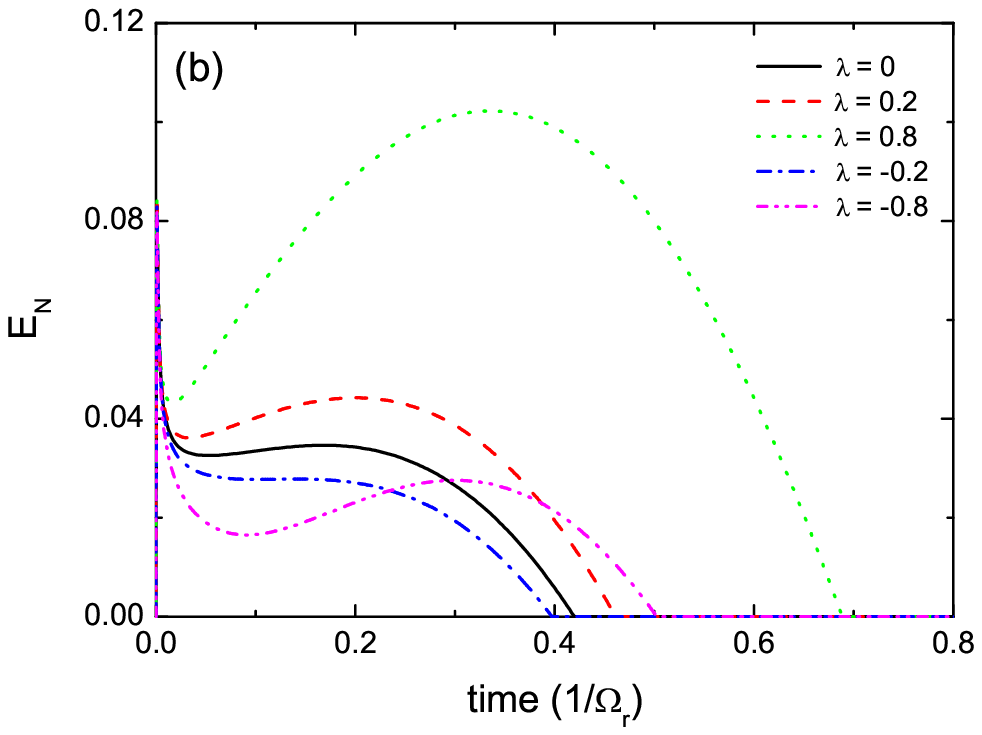}
       \caption{(Color online). Time evolutions of the logarithmic negativity
        for an initial squeezing
        parameter $r=0$. Other conditions and plot caption
        are the same as in Fig.\ \ref{fig_ohmic_LSab1}.
        }\label{fig_ohmic_LSab5}\label{fig_ohmic_LScd5}
  \end{figure}

In Figs.~\ref{fig_ohmic_LSab1} and \ref{fig_ohmic_LScd5}, we plot the
  dynamics of the logarithmic negativity of the two subsystems coupled
to the environments for our
two models with different initial states of $r=2$ and $r=0$, respectively.
Compared with the corresponding $\gamma_0=0$ cases in
Figs.~\ref{fig_iso_squeez2}, 
the oscillatory phenomena due to the influence of the environments
disappear except in Fig.\ \ref{fig_ohmic_LScd1}(b).
It can also be seen from Fig.\ \ref{fig_ohmic_LScd1}(a) that
the entanglement vanishes in finite times (sudden death) \cite{Yu04}.
This is in contrast to the loss of quantum coherence that is usually
  gradual \cite{Yu04,Dodd04,Dodd04a}.

The logarithmic negativity, shown in Fig.\  \ref{fig_ohmic_LSab1}(a)
decays very fast for model A as compared to Model B in Fig.\
\ref{fig_ohmic_LScd1}(b). In other words, the entanglement can
sustain much longer when two
 subsystems are coupled to a common bath than to individually
  independent baths.
This conclusion is consistent with the result found in other
  continuous variable models \cite{Jakub,Benatti06}
or discrete qubit models \cite{Braun02,Benatti03}.
In our models, this could be understood by noting
that the Hamiltonian of the total system can be written in terms of
new dynamical variables, the sum and difference of the two oscillator's
  positions and momenta ($x+y$,
$p_x+p_y$, $x-y$, $p_x-p_y$). For model B coupled to a common
environment, only the mode of the sum of the two positions
 interacts with the environment and
  the mode of the difference of the two positions undergoes a free
  evolution. As a result,
only the modes of the sum of the positions and momenta are affected  by
  the environment
[see Eqs.\ (\ref{master_eqB})]. But for Model A, these modes
all interact with the environments
[see Eqs.\ (\ref{master_eqA})] and thus are all influenced by the
environments.

In Fig. \ref{fig_ohmic_LSab5}(a), we find that the logarithmic
negativity for Model A is barely generated with an initially
separable state ($r=0$ case). 
On the other hand, we find in Fig.
\ref{fig_ohmic_LScd5}(b) that even with no interaction between the
two subsystems, but due to the fact that they coupled to a common
bath (Model B), the entanglement can be generated from an initially
separable state ($r=0$ case)
\cite{Braun02,Benatti03,Jakub,Benatti06}. But the generated
entanglement lasts only for a short time and then disappears. In
most situations the entanglement is created for a very short time
after the interaction with the environment is turned on. The
entanglement may persist for long times or disappear shortly,
depending on system-environment coupling and the properties of the
environment \cite{Braun02,Benatti03,Jakub,Benatti06}.

%%%%%%%%%%%%%%%%%%%%%%%%%%%%%%%%%%%%%%%%%%%%%%%%%%%%%%%%%%%%%%%
\subsection{Condition for Entanglement Survival}
%%%%%%%%%%%%%%%%%%%%%%%%%%%%%%%%%%%%%%%%%%%%%%%%%%%%%%%%%%%%%%%

A condition derived in Ref. \cite{Jakub} stated that if the two-mode
squeezed state of the electromagnetic field embedded in a thermal
environment is initially sufficiently squeezed, it will always
remain entangled independently of the strength of the interaction to
the environment. Each of the two electromagnetic modes has the
Hilbert space equivalent to the Hilbert space of the Harmonic
oscillator. As a result, the model studied in Ref. \cite{Jakub} is
the same as our Model B during $\lambda=0$. However, the conclusion
in Ref. \cite{Jakub} was reached using the RWA-Markovian master
equation. Here we investigate whether the condition presented in
Ref. \cite{Jakub} is still valid or needs some modification in the
non-Markovian case.

In Ref. \cite{Jakub}, the Simon criterion \cite{Simon00} for
continuous variables system was used
  to verify whether the quantum state of the system is entangled or
  separable. It was found that if the initial state
  is sufficiently squeezed, i.e., the squeezed parameter of the initial two-mode squeezed vacuum state satisfies \cite{Jakub},
  \begin{eqnarray}
  \vert r\vert\geqslant\frac{1}{2}\ln(2\bar{N}+1),\label{inequality}
  \end{eqnarray}
  where
  \begin{eqnarray}
  \bar{N}=\frac{1}{e^{\hbar\Omega_r/{k_BT}}-1},\label{nbar1}
  \end{eqnarray}
  is a mean thermal
  photon number, it will remain entangled forever in spite of the
  interaction between the system and the external environment.
  Otherwise, the state will disentangle (become separable)
after time \cite{Jakub}
  \begin{eqnarray}
   t=\frac{1}{2\gamma}\ln\Big(\frac{2\bar{N}+1-e^{-2\vert r\vert}}{2\bar{N}+1-e^{2\vert r\vert}}\Big),\label{critical_time}
  \end{eqnarray}
where $\gamma\approx 2 \lim_{t\rightarrow\infty}\gamma(t)
\approx 2 \gamma_{0}$ 
%is a coupling constant defined in Eq. (\ref{gamma_const}).
From Eqs.~(\ref{inequality}) and (\ref{nbar1}) at a temperature of
  $k_BT=10\hbar\Omega_r$, the corresponding
  critical squeezed parameter
  is $\vert r_c\vert=\frac{1}{2}\ln(2\bar{N}+1)=1.498$.
  We choose the squeezing parameters to be at and slightly smaller
  than this critical squeezing value, and vary the
system-environment interaction strengths to check whether the condition for the
  inequality (\ref{inequality}) is still valid.

  \begin{figure}
    \includegraphics[width=8.7cm]{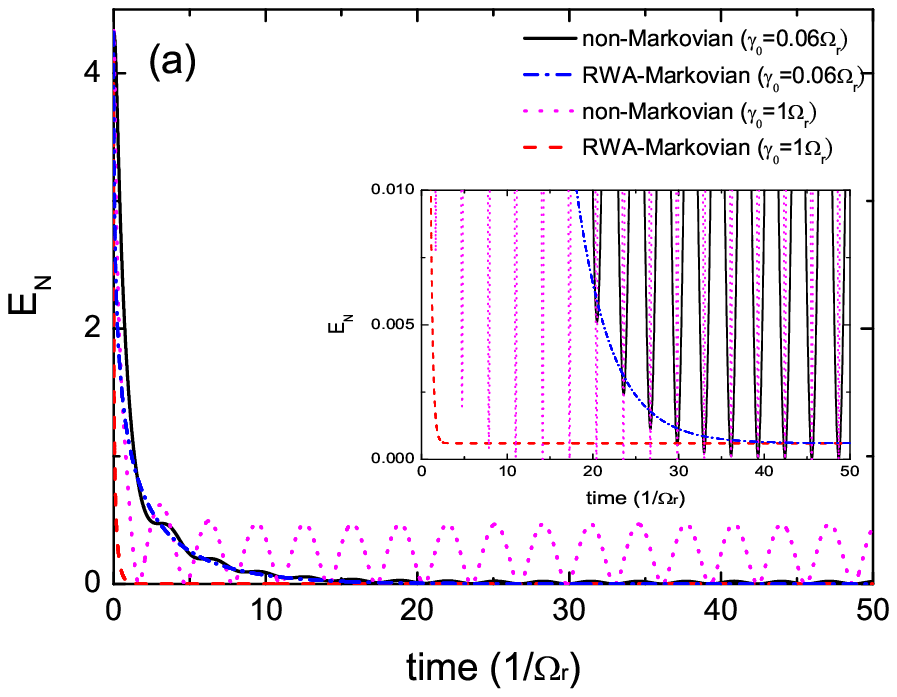}
    \includegraphics[width=8.7cm]{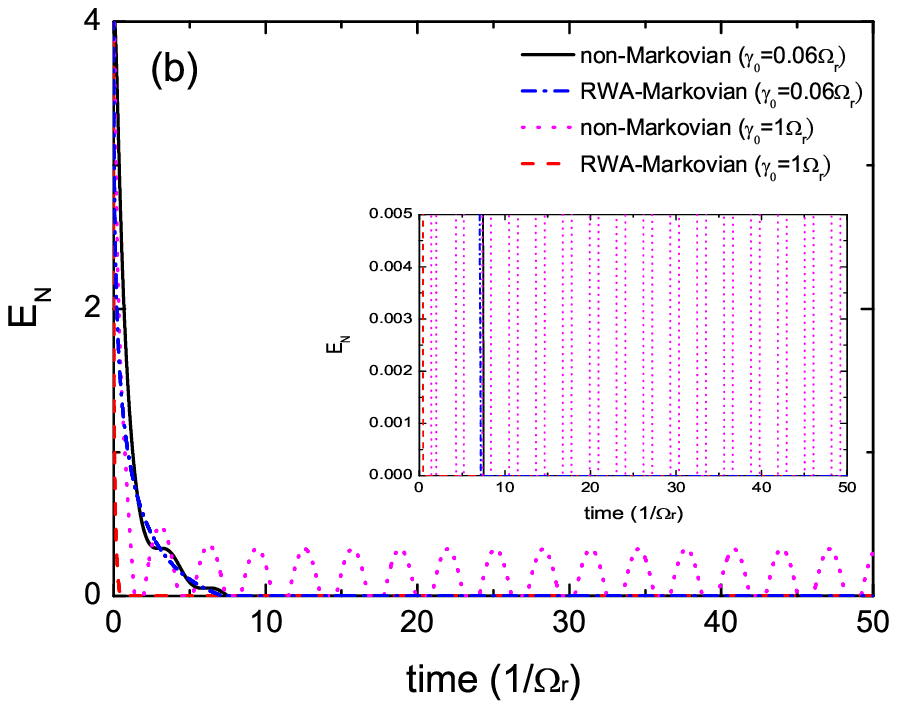}
        \caption{(Color online). The time evolutions of the
    logarithmic negativity for different initial squeezed parameters,
    (a) r=1.489 and (b) r=1.4.
Two different values of $\gamma_0$
($\gamma_0=0.06\Omega_r$ and $\gamma_0=1\Omega_r$ ) are used in each
    plot, where $\gamma_0$ is related to the system-environment
    coupling strength. The insets illustrate the same plots but with
    much smaller values of vertical axis $E_N$.
        }\label{entanglement_existence}
   \end{figure}

  Form Fig. \ref{entanglement_existence}(a) and its inset the logarithmic negativity does not vanish
  and appear cyclically at long time for $r=r_c=1.498$ regardless of their system-environment interaction strength.
  So the statement about the inequality (\ref{inequality}) in
  Ref. \cite{Jakub} seems valid for both non-Markovian and
  RWA-Markovian cases. On the other hand, if the squeezed
  parameter $r=1.4<r_c$, i.e., smaller than the critical squeezed
  parameter, Eq. (\ref{critical_time}) predicts that the two-mode
  state will
  disentangled (or become separable) after time $t=7.168/\Omega_r$ for $\gamma_0=0.06\Omega_r$ and $t=0.43/\Omega_r$
  for $\gamma_0=1\Omega_r$, respectively.
This is indeed the case for RWA-Markovian approximation results shown
  in Fig.~\ref{entanglement_existence}(b) for our Model B with $\lambda=0$.
  However, this is not true for the non-Markovian case.
  We find that entanglement disappear except for the non-Markovian
  case with a larger coupling strength $\gamma_0$, in
  which the entanglement dies out firstly and then be generated
  cyclically by the interaction to the common bath [see
  Fig. \ref{entanglement_existence}(b) and its inset].
In other words, the non-Markovian dynamics predicts that the
entanglement would persist for a longer time. This is consistent
  with the result in Ref.~\cite{Maniscalco07}.
So in the case of non-Markovian dynamics, the inequality
  (\ref{inequality}) and Eq.~(\ref{critical_time}) are no longer true,
  and the condition
  not only depends on the mean thermal photon number
  but also depends on the interaction strength between the system and
  the environment.

%%%%%%%%%%%%%%%%%%%%%%%%%%%%%%%%%%%%%%%%%%%%%%%%%%%%%%%%%%%%%%%
\section{\label{sec6}Conclusion}
%%%%%%%%%%%%%%%%%%%%%%%%%%%%%%%%%%%%%%%%%%%%%%%%%%%%%%%%%%%%%%%

   We have investigated the non-Markovian entanglement dynamics of two
   oscillator subsystems which are 
   coupled to a common environment or are
 coupled respectively to their own independent environments.
 We have presented and discussed the
 influence of the environments on the entanglement dynamics by varying
 initial states (different squeezing parameters), oscillator-oscillator
 interactions and oscillator-environment interactions.
We have found that the dynamics of entanglement oscillates
 for two interacting subsystems isolated from the external environments.
The attractive interaction seems always to be better than the repulsive
interaction as far as the maximum value of entanglement that can be
  reached at a later time is concerned. 
When the coupling between the environments and the two subsystems is turned
   on and
 increased progressively, these periodic behaviors die down gradually and
 disappear eventually. When the interaction strength to the
   environments is increased
   further, the entanglement vanishes at finite times (sudden death).
This is in
   contrast to the loss of quantum coherence that is usually gradual.
It is also been found that the entanglement can sustain much longer
   when the two
 subsystems are coupled to a common bath than to individually
   independent baths.
This conclusion is consistent with the result found in other models
\cite{Jakub}. In summary, the dynamics of the quantum entanglement
is sensitive to the initial states, the oscillator-oscillator
interaction, the oscillator-environment interaction and the coupling
to a common bath or to different, independent baths.

Finally, we have checked the condition for entanglement to exist at
long times for two non-interacting subsystems coupled to a common
bath (model B with $\lambda=0$). In contrast to the condition,
which depends only on the mean thermal phonon number, obtained in
Ref.~\cite{Jakub} using RWA-Markovian master equation, our
non-Markovian analysis indicates that the condition also depends on
the system-environment interaction.

\begin{acknowledgments}
We would like to acknowledge support from the National Science
Council, Taiwan, under Grants No. NSC95-2112-M-002-018 and
No. NSC95-2112-M-002-054.  We also thank support from the focus group
program of the National Center for Theoretical Sciences, Taiwan.
H.S.G. acknowledges support
from the National Taiwan University under Grant No. 95R0034-02, thanks
useful discussions with C.~H.~Chou  and  B.~L.~Hu, and
is grateful to the National Center for High-performance Computing, Taiwan, for
computer time and facilities.
\end{acknowledgments}

% \newpage %Just because of unusual number of tables stacked at end
%\bibliography{mythesis}% Produces the bibliography via BibTeX.
%\begin{bibliography}

\end{document}